\documentclass{article}

\usepackage{cite}
\usepackage{a4wide}
\usepackage{amsmath,amssymb}

\numberwithin{equation}{section}

\DeclareMathOperator{\str}{\mathrm{Str}}

\title{D-Brane Recoil and Supersymmetry Obstruction}
\author{A. Campbell--Smith and N.E. Mavromatos\\{\small \textit{Department of
Physics, King's College London, Strand, London WC2R 2LS}}}
\date{ }

\begin{document}

\maketitle

\begin{abstract}
\vspace{-2in}
\begin{flushright}
hep-th/0003262
\end{flushright}
\vspace{2in}
We discuss a model in which our universe is pictured as a recoiling
Dirichlet brane: we find that a proper treatment of the recoil leads
naturally to supersymmetry obstruction on the four-dimensional world.
An essential feature of our approach is the fact that the underlying
worldsheet sigma model is non-critical, and the Liouville mode plays
the r\^{o}le of the target time.  Also, the extra bulk dimensions are
viewed as sigma model couplings, and as such have to be averaged by
appropriate summation over worldsheet genera.  The recoiling brane is
in an excited state rather than its ground state, to which it relaxes
asymptotically in time, restoring supersymmetry.  We also find that
the excitation energy, which is considered as the observable effective
cosmological `constant' on the brane, is naturally small and can
accommodate upper bounds from observations.
\end{abstract}

\section{Introduction}

In this letter we present a model which describes a four-dimensional
world with \emph{obstructed}~\cite{witten95,acs99:4} supersymmetry:
the vacuum energy vanishes, but the particle spectrum does not respect
supersymmetry.  The four-dimensional world is pictured as a string
soliton (Dirichlet brane) subject to emission/absorption events with
closed string states or other solitonic objects; the resulting recoil
is described~\cite{kogan96:2} by worldsheet logarithmic
operators~\cite{gurarie93,caux96,flohr96,kogan96:1,kogan00},
Liouville-dressed to restore worldsheet conformal
invariance~\cite{david88,distler89}.  The Liouville-dressing furnishes
the model with an extra target-space coordinate, the Liouville mode.
We consider the D-brane soliton embedded in a six-dimensional target
space: the four Neumann coordinates of the D-brane world-volume (with
Euclidean signature), the Liouville mode and one further Dirichlet
coordinate transverse to the brane.  Identifying the Liouville mode
with the target time~\cite{ellis96} and incorporating the effects of
summing over worldsheet genera~\cite{mavro+szabo98} leads to an
induced metric with Minkowskian signature describing the background of
the excited, recoiling brane.  The metric is that of a `thick'
four-dimensional brane-world and is consistent with supersymmetry
obstruction for it has a deficit: this prevents the definition of a
globally covariant spinor supercharge and leads to mass splittings in
the excitation spectrum which do not preserve supersymmetry.  The
`thickness' arises from integrating out the transverse coordinate, as
is appropriate upon its proper identification as a coupling in the
worldsheet sigma model.  This is to be viewed as a dimensional
reduction procedure and a natural method for compactifying extra
transverse dimensions: the dimension is not completely integrated
away, but leads to a `thickened' four-dimensional
world~\cite{lizzi97b}.

We compute the properties of our `thickened' metric from the Einstein
equations (in the presence of a dilaton scalar field related to the
Liouville mode), which proves to be a non-trivial consistency check of
our results.  We find that the five-dimensional cosmological constant
can be made to vanish, consistent with the picture of supersymmetry
obstruction.  This fits naturally with the idea that our world is a
Dirichlet brane which is in an excited state as a result of
recoil~\cite{acs99:4}: asymptotically (in time) supersymmetry in the
mass spectrum will be restored as the brane relaxes.  While the
underlying five-dimensional cosmological constant vanishes, an
observer on the four-dimensional world will measure an effective
cosmological constant arising from the excitation energy of the brane
itself.  We compute the recoil contribution to the excitation energy
of the four-dimensional world and find that it is positive and decays
as $1/t^2$ where $t$ is the target time with origin at the scattering
event causing the recoil.  We also show that a one-loop matter
contribution to the excitation energy on the four-dimensional world
could partially cancel this energy and lead to a naturally small
observed cosmological constant on the brane.

The model proposed here for the appearance of a `thick' brane-world
differs from previously proposed
scenarios~\cite{akama82,rubakov83,visser85,squires86,gogberashvili99,randall99a,randall99b}
in that the `thickening' process appears \emph{dynamically} from a
non-critical (Liouville) string theory.  This latter point of view was
considered in reference~\cite{leontaris99} where, however, the
Liouville mode was not identified with the target time.  In the
current model this identification is crucial for the relaxation
mechanism described above.

\section{Logarithmic operators and D-brane recoil}
\label{sec:recoil}

Consider a string soliton (i.e.\ a Dirichlet brane) with four Neumann
(longitudinal) coordinates $x^\mu$, $\mu \in \{0,1,2,3\}$ and one
Dirichlet (transverse) coordinate $z$.  Notice that from a
target-space point of view a \emph{Euclidean} time $x^0$ is included
in the set of Neumann coordinates.  As we shall see later on a
Minkowskian signature will arise dynamically upon the identification
of $x^0$ with the Liouville mode.  A proper treatment of scattering
events involving either open strings ending on the soliton or closed
string states propagating in the bulk transverse dimension should
include the resulting recoil of the solitonic object.  Note that
qualitatively the effects of these two types of scattering event are
the same, for a closed string state impinging on the soliton will
split into open string states ending on the brane.  The recoil of the
brane under these conditions is known~\cite{kogan96:2} to be described
in terms of a logarithmic conformal field
theory~\cite{gurarie93,caux96,flohr96,kogan96:1} on the open string
worldsheet, and encapsulates the effects of both virtual and on-shell
scattering events.  The pair of logarithmic operators describing the
recoil are
\begin{subequations}\label{logops}
\begin{align}
C^{\mu z} &\doteq \varepsilon g^{\mu z}_C \oint_{\partial \Sigma}
\Theta_\varepsilon(x^\mu) \partial_\perp z, \qquad\text{(no sum on
$\mu$)},\\
D^{\mu z} &\doteq g^{\mu z}_D \oint_{\partial \Sigma} x^\mu
\Theta_\varepsilon(x^\mu) \partial_\perp z, \qquad\text{(no sum on
$\mu$)},
\end{align}
\end{subequations}
where $\partial_\perp$ denotes the normal derivative on the worldsheet
boundary $\partial\Sigma$.  The couplings $g^{ }_C$, $g^{ }_D$ the
recoil/folding of the D-brane soliton as a result of the scattering.
In particular,
\begin{gather}
g_C^{0z} \doteq z_0, \qquad g_D^{0z} \doteq U^z
\end{gather}
where $z_0$ is the origin of the scattering event in the $z$-direction
and $U_z$ is the resulting velocity of the soliton in the
$z$-direction and in general the other components of the two couplings
quantify the folding in the longitudinal space.  In all the above the
parameter $\varepsilon$ is a regulating parameter for the regularized
$\Theta$-function
\begin{gather}
\Theta_\varepsilon (x^\mu) \doteq -i \int d\omega \frac{e^{i\omega
x^\mu}}{\omega - i \varepsilon}.
\end{gather}
This reproduces the normal Heaviside function in the limit
$\varepsilon\rightarrow 0^+$.  To ensure that the pair~\eqref{logops}
satisfies the correct logarithmic algebra one must identify the
regulating parameter with the worldsheet renormalization group scale
\(\ell\doteq\ln|L/a|\) (where $L$ and $a$ are respectively the
infrared and ultraviolet worldsheet configuration space cutoffs)
\begin{gather}\label{epsdef}
\varepsilon^{-2} = 4 \ln \left|\frac{L}{a}\right|.
\end{gather}
As shown in reference~\cite{kogan96:2} the pair~\eqref{logops} are
relevant operators in a worldsheet renormalization group sense with
anomalous dimension $\Delta_C=\Delta_D\doteq\Delta =-\varepsilon^2/2$.
The resulting theory therefore requires Liouville dressing to restore
conformal invariance on the worldsheet~\cite{david88,distler89}.  As
is well known, the Liouville dressing introduces an extra target-space
coordinate $\phi$, the Liouville mode.  The dressed operators are
obtained by first rewriting the boundary operators~\eqref{logops} as
bulk worldsheet operators using Stokes' theorem and then multiplying
the integrand by $\exp{\delta \phi}$ where $\delta$ is the
(worldsheet) gravitational anomalous
dimension~\cite{david88,distler89} which is given by
\begin{gather}\label{deltadef}
\delta = - \frac{Q}{2} + \sqrt{\frac{Q^2}{4} - \Delta} = -\frac{Q}{2}
+ \sqrt{\frac{Q^2}{4} + \frac{\varepsilon^2}{2}}.
\end{gather}
In the above formula $Q^2$ denotes the central charge deficit of the
non-critical theory resulting from the logarithmic
deformations~\eqref{logops}.  On general worldsheet renormalization
group grounds, the central charge deficit $Q^2$ can be computed from
the Zamolodchikov $C$-theorem~\cite{zamolodchikov86} as follows.
Consider the generic worldsheet deformation
\[g^i\int_\Sigma V_i\]
for vertex operators $V_i$ with associated couplings $g^i$.  The
Zamolodchikov metric in this case reads
\begin{gather}\label{zamdef}
{\mathfrak G}_{ij}=\lim_{w\rightarrow 0} |w|^4\langle V_i
(w)V_j(0)\rangle = \delta_{ij} + {\mathcal O}(g^2).
\end{gather}
The $C$-theorem then relates the central charge deficit to the
worldsheet renormalization group $\beta$-functions for the couplings
$g^i$ as follows
\begin{gather}\label{Ctheorem}
Q^2 \sim \int^\ell \beta^i {\mathfrak G}_{ij} \beta^j \sim (\Delta_V
g^i)^2 + {\mathcal O}(g^3).
\end{gather}
It can easily be checked that the second equality in
equation~\eqref{zamdef} is valid also for logarithmic operators
(rewritten as bulk worldsheet operators by means of Stokes' theorem),
as follows directly from their operator product
expansion~\cite{gurarie93,kogan96:2}.

From equation~\eqref{Ctheorem} it is clear that for generic marginally
relevant deformation operators, $Q^2$ is subleading in
equation~\eqref{deltadef}, and as a result the gravitational
anomalous dimension~\eqref{deltadef} is $\delta\simeq\varepsilon$.  We
can confirm this result later as a non-trivial consistency check of
our results.

\section{The Induced Spacetime Metric}
\label{sec:metric}

The Liouville-dressed worldsheet operators in our case read
\begin{subequations}
\begin{align}
C_{\mathrm L}^{\mu z} &\doteq \varepsilon g^{\mu z}_C \int_{\Sigma}
e^{\varepsilon\phi} \partial_\alpha \left(\Theta_\varepsilon(x^\mu)
\partial^\alpha z\right), \qquad\text{(no sum on $\mu$)},\\
D_{\mathrm L}^{\mu z} &\doteq g^{\mu z}_D \int_{\Sigma}
e^{\varepsilon\phi} \partial_\alpha \left( x^\mu
\Theta_\varepsilon(x^\mu) \partial^\alpha z\right), \qquad\text{(no
sum on $\mu$)};
\end{align}
\end{subequations}
here $\alpha\in\{1,2\}$ is a worldsheet index.  By partial integration
these can be re-expressed as a sum of worldsheet bulk and boundary
terms:  the boundary terms describe the recoil excitations in a
conformally invariant fashion, while the bulk operators
\begin{align}
-\int_\Sigma  e^{\varepsilon\phi} \left[ \varepsilon^2 g_C^{\mu z} +
\varepsilon g_D^{\mu z} x^\mu\right]\, \Theta_\varepsilon (x^\mu)\,
\partial_\alpha \phi \partial^\alpha z
\end{align}
describe target-space metric deformation as a result of the recoiling
D-brane~\cite{ellis96}.  The induced metric in the six-dimensional
target space spanned by \(\{\phi,x^\mu,z\}\) and far from and after
the scattering event is
\begin{gather}
G_{\phi\phi} = 1, \quad G_{\phi\mu}=G_{z\mu}=0,\quad
G_{\mu\nu}=-\delta_{\mu\nu}, \quad G_{zz}=-1, \quad G_{\phi z}=
-\sum_\mu \left[ \varepsilon^2 g_C^{\mu z} + \varepsilon g_D^{\mu z}
x^\mu\right].
\end{gather}
Recall that the $g^{ }_{C,D}$ are relevant couplings in a worldsheet sigma
model: they can be made exactly marginal by a
rescaling~\cite{lizzi97a,mavro+szabo98}
\begin{gather}
g^{ }_{C,D} \mapsto \varepsilon \,\bar{g}^{ }_{C,D}
\end{gather}
where $\bar{g}$ is independent of $\varepsilon$.  As shown in
reference~\cite{mavro+szabo98}, summation over worldsheet genera is
equivalent to averaging (with Gaussian weight) over these couplings.
In the limit $\varepsilon\rightarrow 0^+$ the term involving
$\bar{g}^{ }_C$ is subleading and can be dropped.  The averaging
procedure over the remaining coupling is performed around the
classical value $\bar{g}^{ }_D=0$ (corresponding to the classical
static D-brane) and is defined through
\begin{gather} \label{coupavg}
\langle\langle \dots\rangle\rangle \doteq \int d\bar{g}^{ }_D\;
\frac{e^{-\bar{g}_D^2/\Gamma^2}}{\Gamma \surd\pi} (\dots).
\end{gather}
The resulting diagonalized~\cite{leontaris99} averaged line element is
(for $x^\mu$ positive, i.e. away from and after the scattering event)
\begin{align}
\langle\langle ds^2\rangle\rangle &= d\phi^2 - |1-\alpha^2z^2|
\delta_{\mu\nu}dx^\mu dx^\nu - (1+\alpha^2 x_\mu x^\mu)dz^2,\\
\alpha&\doteq \frac{\varepsilon^2\Gamma}{2\surd2} \label{alphadef}
\end{align}
The width $\Gamma$ is known to satisfy a Heisenberg uncertainty
relation in the D-brane recoil picture~\cite{mavro+szabo98}.  In taking
the absolute value in the line element above we imply a mirror
extension around the origin in $z$, which we assume to be the initial
position of the brane.  An important feature of this line element is
the appearance of a singularity at $z=\pm\alpha^{-1}$.  For the
purposes of this letter we will discard the $\alpha^2 x^2$ term in the
fifth component of the metric since this introduces no additional
singular structure and because $\alpha$ is small.

In the spirit of reference~\cite{ellis96} and in contrast to the
approach of reference~\cite{leontaris99} we next identify the
worldsheet zero-mode of the Liouville field $\phi$ with the target
time $x^0\equiv t$.  In spite of the fact that $x^0$ was originally a
Euclidean longitudinal coordinate, after this identification one
obtains a line element with a Minkowskian signature:
\begin{gather}\label{metric}
\langle\langle ds^2\rangle\rangle = \alpha^2 z^2 dt^2 -
|1-\alpha^2z^2|\delta_{ij} dx^i dx^j - dz^2, \qquad i,j\in\{1,2,3\}.
\end{gather}
The identification of the Liouville zero-mode with the target time implies,
on account of equation~\eqref{epsdef}, that $\varepsilon$ scales as
\begin{gather}\label{eptime}
\varepsilon \sim \frac{1}{\surd t}.
\end{gather}
In what follows we shall check the consistency of this identification
by demonstrating that the metric~\eqref{metric} satisfies Einstein's
equations in a five-dimensional target space with a scalar dilaton,
$\varphi$, dependent only on the Liouville
zero-mode~\cite{antoniadis89}:
\begin{gather}\label{dilatondef}
\varphi(t) \doteq \int^t dt' q(t') \doteq M_\mathrm{s}Q(t) t,
\end{gather}
where $M_\mathrm{s}$ is the string scale which we take to be
$10^{19}$~GeV.  In the above, $Q^2(t)$ is the running central charge
deficit of the Liouville theory, whose $t$-dependence arises from
equation~\eqref{eptime}.  This is to be contrasted with standard
non-critical string theory where $Q$ does not
run~\cite{david88,distler89,antoniadis89} and the dilaton has a
component linear in $Qt$.

The satisfaction of Einstein's equations is equivalent to the
restoration of conformal invariance on the worldsheet by the Liouville
dressing procedure~\cite{david88,distler89}.  We will check this
explicitly in section~\ref{sec:eqsM}.  Before doing so, we discuss the
properties of the metric which we have derived.

The $(z,t)$ part of the metric~\eqref{metric} is flat, as can be seen
from the transformations
\begin{gather}
(z,t) \mapsto (u,v)=(z\cosh (\alpha t) ,z\sinh (\alpha t) )
\end{gather}
in which coordinates the metric reads
\begin{gather}
\langle\langle ds^2\rangle\rangle = du^2 -dv^2
-|1-\alpha^2(u^2-v^2)|\delta_{ij}dx^i dx^j.
\end{gather}
This is simply a Rindler wedge (since $u>|v|$) space which is
known~\cite{birrell+davies} to represent (for lines of constant $z$)
the world-line of an accelerated observer with proper acceleration
$1/z$.  However, since our spacetime is only piecewise continuous the
acceleration is only uniform for $t>0$~\cite{ellis97}.  Wedged
spacetimes are known~\cite{witten95} to break supersymmetry because
they obstruct the definition of a global covariantly constant spinor
supercharge.

The structure of this spacetime can be made more transparent by
considering a Euclideanized compactified time coordinate, whence it
becomes clear that the scale $\alpha$ induces a conical deficit.  In
terms of the compactified time coordinate $\tau=2\pi i t/\beta$ where $\tau \in
[0,2\pi)$ and $\beta$ the radius of compactification, the metric reads
\begin{gather}\label{conmetric}
\langle\langle ds^2_{\mathrm E}\rangle\rangle =
-\frac{\alpha^2\beta^2}{4\pi^2} z^2 d\tau^2 - dz^2 -
|1-\alpha^2z^2|\delta_{ij}dx^i dx^j.
\end{gather}
From~\eqref{conmetric} we observe that the deficit is given by
$(2\pi-\alpha\beta/2\pi)$.  A spacetime with such a conical deficit is
known to exhibit \emph{supersymmetry
obstruction}~\cite{witten95,acs99:4} in which, although the vacuum
state energy remains vanishing, supersymmetry is obstructed in the
spectrum of massive excitations by splittings $\delta m \propto
\alpha$.  Equivalently we can view this spacetime as being in thermal
equilibrium with a bath of `temperature' $T=\beta^{-1}=\alpha/2\pi$~\cite{unruh76} in
which case the deficit disappears but the presence of a non-zero
temperature gives rise to non-supersymmetric mass splittings in the
excitation spectrum $\delta m \propto T$ as before~\cite{derendinger98}.

The `thickened' brane-world picture is obtained now by considering the
effects of a proper sum over worldsheet genera~\cite{mavro+szabo98}
which implies that the Dirichlet coordinate $z$, as a coupling in a
worldsheet sigma model, should also be averaged in analogy with
equation~\eqref{coupavg}.  The Gaussian nature of this averaging
implies that averages of odd powers of $z$ will vanish.  This picture
will be incorporated at the level of a target-space effective action
in the next section by considering a `stack' of parallel D-branes in
the $z$-direction~\cite{lizzi97b} each with their own excitation
energy from the recoil.  This `stacking' is a purely formal
construction descibing a single physical brane fluctuating around
$z=0$ and should not be confused with the matrix D-brane structures of
reference~\cite{mavro+szabo98} which involve open strings stretching
between the branes.  The `stack' of D-branes here implies an explicit
breaking of five-dimensional Lorentz symmetry; the four-dimensional
subgroup of Lorentz rotations on the brane remains unbroken.

\section{The Dynamics}
\label{sec:eqsM}

From a standard low-energy sigma model point of view~\cite{metsaev87}
the gravitational part of the (order $\alpha'$) effective action in
our five-dimensional spacetime is given by
\begin{gather}\label{effac}
S = M^3_\mathrm{s}\int d^5x\; \sqrt{-g}\, e^{-\gamma\varphi} \left[
\mathcal{R} + \beta (\nabla \varphi)^2 -\Lambda \right] -
M^2_\mathrm{s}\sum_i \int d^4x\; \sqrt{-g^{(4)}(x,z_i)}
e^{-\gamma\varphi} V(x,z_i).
\end{gather}
The interval of the $z$-integration is between the singularities at
$\pm \alpha^{-1}$.  For convenience we have absorbed factors of
$M^{-2}_\mathrm{s}$ into both $\Lambda$ and $V$.  The Einstein frame
corresponds to $\gamma=0$ and $\beta<0$.  On the other hand, in the
sigma-model frame~\cite{metsaev87} $\gamma>0$ and $\beta>0$.  As
appropriate for a Liouville string
theory~\cite{david88,distler89,antoniadis89}, in what follows we shall
work exclusively in the sigma-model frame.  The sum in the last term
in the action~\eqref{effac} describes the `stack' of D-branes in the
$z$-direction, each of which has a possible recoil excitation energy
$V(x,z_i)$.  This should be contrasted with the orbifold construction
of reference~\cite{randall99a} where there are only two D-branes.  In
our case there is a continuum representation of this sum in terms of
an averaged `thickened' D-brane~\cite{lizzi97b}:
\begin{gather}\label{contint}
\sum_i \int d^4x\; \sqrt{-g^{(4)}(x,z_i)} e^{-\gamma\varphi} V(x,z_i)
\longrightarrow M_\mathrm{s}\int d^5x\; \sqrt{-g^{(4)}(x,z)} e^{-\gamma\varphi}
V(x,z);
\end{gather}
the continuum $V(x,z)$ incorporates any non-trivial $z$-dependent
measure appearing in the passage to the continuous form and the
function $V$ denotes the effective recoil excitation energy as
measured on the four-dimensional brane-world.  Notice, however, that
as a result of the `stack' of D-branes the five-dimensional integral
in equation~\eqref{contint} does not contain a dependence on the fifth
component of the metric.  This means that upon a variation with
respect to the fifth component of the metric, this term will not
contribute.  With this in mind the equations of motion derived from
the effective action~\eqref{effac},~\eqref{contint} are
\begin{subequations}\label{einsteqs}
\begin{align}
&R_{\mu\nu}-\frac{1}{2}g_{\mu\nu}{\mathcal R} = -\frac{1}{2}g_{\mu\nu}
[\Lambda+V] -\partial_\mu \varphi \partial_\nu \varphi + \frac{1}{2}
g_{\mu\nu} (\partial \varphi)^2, \qquad \mu,\nu \in \{ 0,1,2,3\},\\
&R_{55}-\frac{1}{2}g_{55}{\mathcal R} = -\frac{1}{2}g_{55}
[\Lambda] -\partial_5 \varphi \partial_5 \varphi + \frac{1}{2}
g_{55} (\partial \varphi)^2,\label{fiftheinst}
\intertext{and for the dilaton}
&\mathcal{R} -[\Lambda+V] +2(\nabla\varphi)^2
-2\nabla^2\varphi =0.\label{dileqM}
\end{align}
\end{subequations}

As mentioned in the previous section, summation over worldsheet genera
requires an averaging procedure for the Dirichlet coordinate:
$\langle\langle z^2\rangle\rangle_z = \sigma^2$, which leads to the
metric (recalling the identification~\eqref{eptime})
\begin{align}\label{lastmetric}
\langle\langle ds^2\rangle\rangle &= \frac{b^2\sigma^2}{t^2} dt^2 -
\left|1-\frac{b^2\sigma^2}{t^2}\right| \delta_{ij}dx^i dx^j -dz^2,\\
b&\doteq \frac{\Gamma}{2\surd2}.
\end{align}
For notational convenience we have used the symbol
$\langle\langle\dots\rangle\rangle$ to denote the combined Gaussian
average over both logarithmic couplings $\bar{g}^{ }_C$ and $\bar{g}^{
}_D$.  The components of the Ricci tensor and curvature scalar for
this metric read
\begin{align}
R_{00} &= \frac{3b^2\sigma^2}{\left|1-\frac{b^2\sigma^2}{t^2}\right|^2 t^6}
\left[ (b^2-2)t^2 + b^2\sigma^2(1-b^2) \right],\\
R_{ij} &= \delta_{ij}
\frac{(b^2-1)}{\left|1-\frac{b^2\sigma^2}{t^2}\right|t^4} \left[ 3b^2\sigma^2 -
2t^2\right],\qquad i,j \in\{1,2,3\},\\
R_{55} &= -\frac{3b^2}{\left|1-\frac{b^2\sigma^2}{t^2}\right|^2t^2},\\[.5cm]
\mathcal{R} &= \frac{12(b^2-1)}{\left|1-\frac{b^2\sigma^2}{t^2}\right|t^2}.
\end{align}
We stress that the above components are computed from the averaged
metric~\eqref{lastmetric} and are not simply averaged versions of
equations~\eqref{einsteqs}.  This is essential for our dimensional
reduction proposal, which should be viewed as an alternative to
compactification: in our approach the fluctuations of the graviton
field on the brane are $z$-independent as a result of the above
average over the Dirichlet coordinate/coupling $z$.  As a result,
Newton's law on the four-dimensional world is unaffected in contrast
to other proposals where $z$ remains a fully-fledged
coordinate~\cite{randall99a,randall99b,floratos99,brandhuber99,leontaris99,csaki00}.
However, in our picture the effect of this Dirichlet coordinate is
still felt in the `thickening' of the recoiling brane and in the
constraint coming from the fifth component of the Einstein
equation~\eqref{fiftheinst}.  It is understood that the above picture
pertains strictly to the case of a stack of identical, parallel
solitons; for the intersecting-brane case the r\^{o}le of the
Dirichlet and Neumann coordinates are mixed and our construction
fails.

To leading order in $t\ggg1$ the solutions of the Einstein
equations~\eqref{einsteqs} read
\begin{align}\label{qeqn}
q^2(t) &= \frac{b^2\sigma^2(4-b^2)}{t^2} \; >0 \quad\text{for}\quad b^2<4,\\
\Lambda &= \frac{(5b^2-8)}{t^2},\qquad\qquad
V= \frac{(2b^2+4)}{t^2} \; >0.\label{Vrecoil}
\end{align}
From equations~\eqref{qeqn} and~\eqref{dilatondef} we obtain the
central charge deficit
\begin{gather}
Q^2(t) = \frac{b^2\sigma^2(4-b^2) (\ln t)^2}{t^2}
\end{gather}
so for $b^2<4$ the worldsheet sigma model is
supercritical~\cite{antoniadis89} and our identification of the
Liouville mode with the target time is self-consistent.  As
anticipated in section~\ref{sec:recoil}, the central charge deficit is
indeed subleading in equation~\eqref{deltadef}, $Q^2\sim
\mathcal{O}(\varepsilon^4 (\ln\varepsilon)^2)$.

In the classical limit, $\sigma\rightarrow 0$ the dilaton equation of
motion implies $\mathcal{R} -\Lambda -V=0$ which puts a dynamical
constraint on the width parameter, $b^2=8/5$.  This constrains the
cosmological term to vanish for a value of $b$ compatible with the
supercriticality of the worldsheet sigma model.  If one demands
continuity with the quantum case where $\sigma$ is identified with the
position uncertainty of the brane then we observe that this constraint
can be preserved by the very natural choice $\sigma=
5/\sqrt{96}M_\mathrm{s} \simeq 1/2M_\mathrm{s}$ compatible with
near-saturation of the position-momentum uncertainty relation for
D-branes~\cite{mavro+szabo98}.  Note also that, as expected, the
recoil excitation energy $V$ is positive definite and that both $Q(t)$
and $V(t)$ relax so that asymptotically criticality and supersymmetry
are restored~\cite{acs99:4}.

\section{One-Loop Matter Contribution to the Excitation Energy}
\label{oneloopV}

In our picture all the matter fields on the four-dimensional
brane-world are viewed as excitations on the already excited (recoiling)
D-brane.  We can estimate their contribution to the excitation energy
in our picture from the one-loop effective potential used to calculate
the vacuum energy in standard four-dimensional (supersymmetric) field
theory~\cite{coleman73,weinberg73,iliopoulos75,ferrara94}:
\begin{gather}
V_1 \ni \frac{M_{\mathrm{uv}}^2}{32\pi^2} \str M^2
\end{gather}
where $M_{\mathrm{uv}}$ is the ultraviolet cutoff and 
\begin{gather*}
\str M^n = \sum_i (-1)^{2J_i} (2J_i+1) M_i^n,
\end{gather*}
whence
\begin{gather}\label{strace}
\str M^2 \sim 2(M^2_{\mathrm b} - M^2_{\mathrm f}).
\end{gather}
We noted in section~\ref{sec:metric} that the metric describing the
background recoiling brane can be interpreted in terms of
thermalization with temperature $T=\alpha/2\pi$.  With this
interpretation the `thermal' mass splittings required to evaluate the
supertrace~\eqref{strace} can be computed from a thermal superspace
formalism: the thermal mass modes for bosonic and fermionic
excitations are respectively given by~\cite{derendinger98}
\begin{align*}
M^2_{\mathrm b} &= M^2_{\mathfrak L} + 4\pi^2 n^2 T^2, \qquad n\in\mathbb{Z},\\
M^2_{\mathrm f} &= M^2_{\mathfrak L} + (2n+1)^2\pi^2 T^2, 
\qquad n\in\mathbb{Z}.
\end{align*}
In the above, $M_{\mathfrak L}$ is the mass term which appears in the
zero temperature Lagrangian density.  Now the supertrace can be
computed by summing~\eqref{strace} over the modes $n$; regularizing
the divergent sums using $\zeta$-function regularization the result is
a sum over species living on the four-dimensional world:
\begin{gather}
V_1 \ni - \sum_{i=1}^{N_\mathrm{s}}\frac{b^2}{t^2}
\frac{M^2_\mathrm{uv}}{8\pi^2} \zeta(-1,{\textstyle
\frac{1}{4}})
= \frac{b^2}{t^2}\frac{N_\mathrm{s}M^2_\mathrm{uv}}{8\pi^2}
\frac{B_2({\textstyle \frac{1}{4}})}{2}
= -\frac{N_\mathrm{s}M^2_\mathrm{uv}b^2}{384\pi^2 t^2},
\end{gather}
where $B_2(x)=x^2-x+\frac{1}{6}$ is the second Bernoulli
polynomial~\cite{gradshteyn+ryzhik}.  Comparison with the recoil
contribution to $V$ in equation~\eqref{Vrecoil}, shows that there is
the possibility to achieve a naturally small total excitation energy
$V$ if the ultraviolet cutoff has the not unnatural value
\begin{gather}
M_\mathrm{uv} \sim \frac{20\pi}{\surd N_\mathrm{s}} M_\mathrm{s}.
\end{gather}
Note that in the Standard Model $N_\mathrm{s}\sim40$.  The resulting
suppression allows us to choose the supersymmetry obstruction scale
$\alpha$ in~\eqref{alphadef} to be of the order of a few TeV without
generating an unacceptably large observable cosmological `constant'
(the r\^{o}le played by $V$) on the four-dimensional brane world,
compatible with the astrophysical upper bound of $10^{-120}$ in Planck
units.  Note that global supersymmetry obstructed at scales of a few
TeV provides a solution to the gauge hierarchy problem stable to
higher order quantum corrections.  In the case of supergravity
theories, which describe the physics on brane-worlds, this conclusion
is not always valid with the notable exceptions of effective theories
derived from string models~\cite{ferrara94}.  A detailed study of such
issues in the context of the present model is beyond the scope of the
current work.

\section{Discussion}

In this letter we have presented a mechanism whereby supersymmetry
obstruction is realized dynamically in a recoiling D-brane framework.
An essential feature of our approach is that dimensional reduction to
four spacetime dimensions occurs as a result of an averaging procedure
over Dirichlet coordinates in a worldsheet sigma model.  As a result
of this averaging procedure, the four-dimensional brane-world
metric~\eqref{lastmetric} can be recast as a standard
Friedmann--Robertson--Walker metric by the rescaling $t\mapsto
t_\mathrm{FRW}= b\sigma\ln t$.  In this frame the excitation energy
$V$, the observable cosmological `constant' on the four-dimensional
world, relaxes as $\exp\{-2t_\mathrm{FRW}/b\sigma\}$ and the
dilaton~\eqref{dilatondef} is linear in $t_\mathrm{FRW}$: $\varphi =
12t_\mathrm{FRW}/5$~\cite{antoniadis89}.  Phenomenologically, in this
framework, to ensure that the supersymmetry obstruction scale is of
order a few TeV, one needs recoil events with temporal separation
$t_\mathrm{FRW}\sim20M^{-1}_\mathrm{s}$, which is a natural timescale
for quantum gravity effects, and therefore consistent with the
spacetime foam picture resulting from recoiling
D-branes~\cite{ellis99}.

\subsection*{Acknowledgements}

The authors would like to thank John Ellis, D.~V. Nanopoulos,
K. Tamvakis and J.~F.  Wheater for valuable discussions.  A.C.--S. and
N.E.M. acknowledge partial financial support from the Leverhulme Trust
and P.P.A.R.C. (U.K.)  respectively.

\end{document}